\documentclass[twocolumn,showpacs,preprintnumbers,amsmath,amssymb]{revtex4}

\usepackage{graphicx}
\usepackage{dcolumn}
\usepackage{bm}
\usepackage{epsfig}

\begin{document}

\title{Extension to the Quantum Langevin Equation in the Incoherent
  Hopping Regime.} 
\author{Andrew G. Green}
\affiliation{School of Physics and Astronomy, University
 of St Andrews, North Haugh, St Andrews, Fife, KY16 9XP, UK.}
\email{andrew.green@st-andrews.ac.uk}
\date{\today}
\begin{abstract}
An extension to the quantum Langevin equation is derived, that is valid in the
incoherent hopping regime, and which allows one to incorporate quantum
tunneling events. This is achieved by the
inclusion of additional stochastic variables in the Langevin
equation representing the tunneling events. A systematic
derivation of this extension and of its regime of validity is
presented. The study 
is motivated by efforts to determine the error in reading the state of
a super-conducting quantum bit. 
\end{abstract}
\maketitle
\begin{narrowtext}

The quantum Langevin equation\cite{Schmid82,Eckern84} provides a
physically appealing and numerically  simple description of 
the dynamics of a system coupled to a heat bath. It 
describes the system's non-linear semiclasical dynamics and 
the effects of coupling to quantum and thermal fluctuations in the heat
bath. It does not, however,  describe quantum tunneling of the system. 
Instanton approaches\cite{Ford88,Ivlev85,Fisher88} provide a 
powerful framework in which to describe these latter effects.
Here, 
we synthesize
these two methods within an
extended  quantum Langevin equation that includes
quantum tunneling. 
We focus upon two effects; quantum tunneling through
or reflection from an energy barrier. These may be incorporated 
into the Langevin equation by the inclusion of additional stochastic
variables, representing the  
tunneling and reflection events. This is valid in the regime of
incoherent hopping.

Specifically, 
we concentrate upon the case of the Josephson
junction. Quantum Langevin equations have been used extensively in the
study of such systems. Recent
experiments\cite{Vion02,Devoret} have probed Josephson junctions in
regimes where current analytical tools do not provide a full
description. In the experiments of Refs.\cite{Vion02,Devoret}, a
Josephson junction was used to read out the state of a charge/phase
quantum bit. 
%
The most recent\cite{Devoret} made novel use of the {\it
  non-linear}, semi-classical dynamics of the junction through its
vicinity to a classical bifurcation instability. Readout
incurred errors due to quantum effects
(tunneling between phase-space
trajectories\cite{Dykman}) and thermal and quantum fluctuations in the
transmission line.
Here, we consider an experimentally realisable alternative scheme where the
junction absorbs energy from a chirped microwave pulse with frequency
modulated to match the response of the junction (whose
frequency falls as the amplitude of its response increases). Depending
upon the state of the qubit, the junction will either absorb enough
energy to surmount the barrier--- leading to a $\pi$ phase-shift in the
reflected signal--- or not. The readout incurs errors due to the effect
of the bath and due to quantum tunneling/reflection at the barrier.

In order to fully understand such systems one must describe
 the non-linear semi-classical dynamics and quantum tunneling 
in concert with coupling to thermal and quantum fluctuations
 in the bath. 
One approach is to follow the evolution of the density matrix through
 its master equation. 
This is numerically extremely intensive when the
Josephson junction contains more than a few levels.
An 
extension of the
 Langevin equation is, therefore, highly desirable\cite{Footnote1}. We
first give a heuristic   
justification for such an extension, before turning to a derivation
 within a Keldysh field theory. 
We describe how this extended Langevin equation may be used to estimate
 errors in the chirped pulse readout scheme.

Consider a Josephson junction coupled by a transmission lead of
impedance $Z$ to measurement  apparatus. The quantum Langevin equation
is given by\cite{Ford88}
\begin{eqnarray}
& &
\left(\frac{\hbar}{2e} \right) C_J \ddot \phi
+I_J\sin \phi(t)
+ \frac{1}{Z}\left(\frac{\hbar}{2e} \right) \dot \phi(t)
=
I_N(t)+I_D(t)
\nonumber\\
& &
\langle I_N(\omega) I_N(-\omega) \rangle
=
2 \hbar \omega \frac{1}{Z} \coth[ \hbar \omega/2T],
\label{QLangevin1}
\end{eqnarray}
$\phi(t)$ is the phase difference of the superconducting order
parameter across the junction, $C_J$ is the 
junction capacitance
and $I_J$ is the Josephson current 
($I_J=(2e/ \hbar)E_J$, where $E_J$ is the Josephson energy).
Fluctuations in the transmission line produce a current noise $I_N(t)$.
$I_D(t)$ is a drive current produced by the microwave source. 

It is useful to explicitly introduce the charge, $n(t)$, on the
junction and re-write the quantum Langevin equation (\ref{QLangevin1})
as a pair of coupled equations;
\begin{eqnarray}
\left(\frac{\hbar}{2e}\right) \dot \phi- \frac{n}{C_J}
=0
\nonumber\\
\dot n - \left(\frac{\hbar}{2e}\right) \frac{\dot \phi}{Z} + \dot q +
I_J \sin(\phi) =I_D(t)
\nonumber\\
\langle q(\omega) q(-\omega) \rangle
=\frac{2\hbar }{\omega Z} \coth(\hbar \omega/2T)
\label{QLangevin2}
\end{eqnarray}
The noise is now given in terms of fluctuations in the charge on the
transmission lead, $q(t)$. Eliminating $n$ from Eqs.(\ref{QLangevin2})
and substituting $\dot q=I_N$ recovers Eq.(\ref{QLangevin1}). 

The dynamics of the Josephson junction is 
equivalent to that of a rigid pendulum; the superconducting phase
difference plays the role of the angle of the pendulum and the
Josephson potential the role of the gravitational potential. 
%
%
If the pendulum does not quite have
enough energy to perform a complete revolution, classically, it will
momentarily stop at some angle short of the vertical and then reverse
its motion. Quantum mechanics allows the pendulum to
tunnel through the remaining potential
barrier. 
If the pendulum does have sufficient energy to pass through the
vertical, the pendulum may be quantum mechanically reflected from the
barrier and its motion reversed.
These two processes can be represented by boundary
conditions for the phase of the junction;
$
\phi \rightarrow -\phi 
$
at 
$
t(\dot \phi=0)
$
and
$
\dot \phi  \rightarrow -\dot \phi 
$
at
$
t(\phi=\pm \pi)
$
or, re-expressing these using the first of Eqs.~(\ref{QLangevin2})
\begin{eqnarray}
\phi &\rightarrow& -\phi 
\;\;\;\;\;
\hbox{ at} 
\;\;\;
t(n =0)
\nonumber\\
n & \rightarrow& -n 
\;\;\;\;\;
\hbox{ at} 
\;\;\;
t(\phi=\pm \pi)
\label{Boundary_Conditions}
\end{eqnarray}
At each juncture where the junction/pendulum
is momentarily stationary short of the vertical or passing through
the upwards vertical, with some
probability, $\phi$  and $n$ are modified by quantum
processes according to Eqs.~(\ref{Boundary_Conditions}).
 We may heuristically modify the
Langevin equation in order to incorporate these effects by the
addition of suitable $\delta$-functions as follows:
\begin{eqnarray}
& &
\left(\frac{\hbar}{2e}\right) \dot \phi- \frac{n}{C}
=-2 \left(\frac{\hbar}{2e}\right) \sum_a \eta_a \phi(t_a) \delta(t-t_a)
\nonumber\\
& &
\dot n - \left(\frac{\hbar}{2e}\right) \frac{\dot \phi}{Z} + \dot q +
I_J \sin \phi
\nonumber\\
& &
\;\;\;\;\;\;\;\;\;\;\;\;\;\;\;\;\;\;\;\;\;\;\;\;\;=
-2  \sum_b \eta_b n(t_b)\delta(t-t_b) +I_D(t)
\nonumber\\
& &
\langle q(\omega) q(-\omega) \rangle
=\frac{2\hbar}{\omega Z} \coth(\omega/2T).
\label{Extended_Langevin}
\end{eqnarray}
$t_a$ are times such that $n(t_a)=0$ and $t_b$ such
that $\phi(t_b)= \pm \pi$. $\eta_a$ and $\eta_b$ are additional
stochastic variables accounting for the probability of quantum
tunneling/reflection. They take the values $\eta=0$ or $1$ with a
probability $P(\eta)$ that is a function of the energy of the
junction at the stationary point, $E=E_J(1+\cos
\phi(t))+n^2/2C$. Without a bath, 
$P(\eta=1)=1/(e^{2\pi E/ \Omega}+ 1)$, where 
$\Omega=(2e/\hbar) \sqrt{E_J/C} $
is the curvature at the top of the junction
potential. Coupling to the bath renormalizes the potential\cite{Ford88} giving
$\Omega=(2e/\hbar) \sqrt{E_J/C} 
\left[1-  (2e/\hbar) \sqrt{E_J/Z^2C^3}\right]$. 
Further corrections occur due
to the finite drive\cite{Ivlev85,Fisher88} and may be
incorporated {\it via} additional modifications to $P(\eta)$. In the case of
the chirped drive, the potential is zero at the stationary points and these
corrections are not required.

In a chirped readout, the junction will gradually absorb energy from
the microwave drive and go through a number of stationary points at
which it approaches within an (decreasing) energy $E$ of the top
of the junction potential. Coupling to fluctuations in the transmission
line leads to an energy uncertainty $\Delta E=T$ at high temperatures;
this may be seen by linearizing Eq.(4) around the classical,
zero-noise solution with a chirped drive. The error probability at
each stationary point is given by the convolution of the distribution
of arrival energy with the tunneling probability. The total error
probability is given by 
$\sum_{n=1}^{n=n_{max}} [\prod_{i=1}^{n-1}(1-p_i)]p_n $,
where $p_n$ is the error probability at the nth stationary
point. When the chirped drive takes only a few cycles to bring the
junction to the top of its potential, only the error probability
$p_{n_{max}}$ will be significant. In this case, analytic
approximations to the error probability may be made. In the more
general case, numerical integration of Eq.(4) provides an elegant way
of calculating the cumulative effect of errors.


Next, we derive the modified Langevin Eq.(\ref{Extended_Langevin})
from a Keldysh field theory. 
The kernel for propagation of the density matrix may be
expressed as a Schwinger-Keldysh field
theory\cite{Schwinger61,Keldysh64,Feynman63,Rammer86}. In essence, the
Keldysh field theory expresses the propagation of the right and left
projections of the density matrix in terms of fields which propagate forwards
and backwards in time\cite{Rammer86}.
It is useful to make a change of basis to the sums and differences of
these fields. Since classical field configurations take the same value
for both forwards and backwards propagation in time, the sum is known
as the classical component (denoted below with a subscript $c$) and the
difference is known as the quantum component (denoted with a subscript
$q$). In certain circumstances (discussed below) the effective
theory of the classical components can be expressed as a quantum
Langevin equation\cite{Schmid82}.

The Keldysh Lagrangian for the Josephson junction is
\begin{eqnarray}
& & {\cal L}(n,\phi,Q,V)
\nonumber\\
&=&
\frac{\hbar}{2e} n \sigma^{qc}_x\dot \phi 
-\frac{n \sigma_x^{qc}n}{2C_J}
+E_J \left(\cos(\phi_c+\phi_q)-\cos(\phi_c-\phi_q) \right)
\nonumber\\
& &
+Q \sigma_x^{qc} \left(\frac{\hbar}{2e} \dot \phi - V\right)
+ \frac{1}{2} V(t) \int_{-\infty}^{\infty} dt'{\bf D}^{-1}(t,t') V(t').
\label{Keldysh_Lagrangian}
\end{eqnarray} 
The fields $n$, $\phi$, $Q$ and $V$ are vectors in
the Keldysh (quantum/classical) space, {\it e.g} $\phi \equiv
(\phi_c(t), \phi_q(t))$, except where the
indices $q$ or $c$ are given explicitly. $Q$ is a Lagrange
multiplier that imposes the Josephson relation $V=\hbar \dot \phi/2e$
and $V$ is the voltage across the junction. The first term expresses
the fact that $\phi$ and $n$ are
conjugate, the second term gives the charging energy and the third
term the Josephson energy (or non-linear inductance energy). The final
term describes voltage fluctuations at the end of the
transmission line. The correlation matrix is given, in the Keldysh basis, by
\begin{eqnarray*}
{\bf D}(t_1,t_2)
=
-i \langle V_{\alpha}(t_1) V_{\beta}(t_2) \rangle
=
\left(
\begin{array}{cc}
D_K(t_{12}) & D_R(t_{12}) \\
D_A(t_{12}) & 0
\end{array}
\right).
\end{eqnarray*}
where the subscripts $K$, $R$ and $A$ refer to Keldysh, retarded and
advanced correlators, respectively. These
are calculated for a transmission lead in thermal equilibrium;
\begin{eqnarray*}
D_R(t_{12})
&=&
D_A(t_{21})
=
-\frac{i}{2} \theta(t_1-t_2)\langle [V(t_1),V(t_2)] \rangle
\nonumber\\
D_K(t_{12})
&=&
-\frac{i}{2} \langle \{ V(t_1),V(t_2) \} \rangle 
\end{eqnarray*}
Their Fourier transforms are $D_R(\omega)=i(\omega-i \delta) Z$ and 
$D_K(\omega)=2i \omega Z \coth (\omega/2T)$. 
The quantum Langevin equation (\ref{QLangevin2}) can be derived from
Eq.~(\ref{Keldysh_Lagrangian}) as follows\cite{Schmid82,Eckern84}: 
After integrating over $Q$,
$V$ is replaced by $\hbar \dot \phi/2e$ in the final term. 
The
Josephson potential term is linearized in $\phi_q$, reducing it to
$ 2 \phi_q E_J \sin \phi_c$. This
assumes that the quantum mechanical spread of the phase difference is
small\cite{Aashish}. 
The term quadratic in $\phi_q$
is decoupled {\it via} a Hubbard-Stratonovich transformation with a
field $\dot q$. 
The resulting theory is linear in $\phi_q$ and
$n_q$. Integrating over these fields leads to
$\delta$-functionals that impose Eqs.(\ref{QLangevin2}). 
The remaining quadratic Lagrangian for $q(t)$
determines its correlation function; 
$\langle q(\omega) q(-\omega) \rangle =D_K(\omega)/[D_R(\omega)
D_A(\omega)]$.
After these manipulations, the Keldysh partition function
reduces to 
\begin{eqnarray}
{\cal Z}
&=&
\int D\phi_c D n_c
\delta 
\left[ 
\left(\frac{\hbar}{2e}\right) \dot \phi- \frac{n}{C} 
\right]
\nonumber\\
& &
\times
\delta
\left[
\dot n - \left(\frac{\hbar}{2e}\right) \frac{\dot \phi}{Z} + \dot q +2
I_J \sin(\phi)
\right]
\nonumber\\
& &
\times
\exp \left[ -
\frac{\hbar }{\omega Z} \coth(\hbar \omega/2T)q(\omega)
q(-\omega)
\right]
\label{Partition1}
\end{eqnarray}
which is equivalent to Eqs.(\ref{QLangevin2}).

This derivation ignores instantons; imaginary-time diversions of the
Keldysh contour. These describe the quantum tunneling and reflection
discussed in our heuristic derivation of the extended
Langevin Eq.(\ref{Extended_Langevin}). 
Instantons occur on both the outgoing and return parts of the Keldysh
contour, as shown in Fig.1. Each dot corresponds
to an imaginary time excursion in the evolution of the fields in the
path integral. During the imaginary-time evolution, the potential is
inverted and the equations of motion are such that the particle can
travel between minima of the potential in a classically 
forbidden way\cite{Rajaraman}. 

Consider the effect of a single instanton at time $t_0$ on the upper
part of the Keldysh contour that
takes the forward propagating $\phi_+$ to $ -\phi_+$. 
This instanton leads to boundary conditions
$\phi_+(t_0-0^+)=-\phi_+(t_0+0^+)$ in the path integral for the
real-time evolution of the fields $\phi(t)$ and $n(t)$.
The probability of this
instanton is given by the exponential of its
classical action\cite{Rajaraman}, possibly modified by finite drive
currents\cite{Ivlev85,Fisher88}. The full
path 
integral includes a sum over all possible configurations and numbers
of instantons, weighted with the appropriate probability. 
In the absence of a bath, these
instantons are positioned independently on the Keldysh contour each
occurring at a stationary point of the classical
evolution\cite{Rajaraman}.
 
\begin{figure}[htbp]
\begin{center}
\epsfig{file=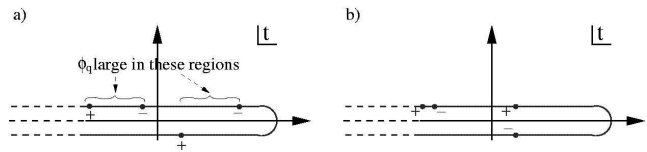, height=0.75in}
\caption{{\bf Schematic Diagram of the Real Time Integration Contour
    Including Instantons.}  
a) Shows the regime of coherent hopping where instantons are unbound. 
b) Shows the regime of incoherent hopping, where instantons are bound
in pairs.} 
\label{Instantons}
\end{center}
\end{figure}

Coupling to bath fluctuations induces interactions
between instantons, tending to bind them in pairs. Pairing occurs both
between instantons 
and anti-instantons on the same part of the Keldysh contour and
on opposite parts of the the Keldysh contour.
Consider 
one instanton and one anti-instanton both on the upper contour. In the
time between these two instantons, the fields on the upper and
lower contour are very different and $\phi_q$ is large. A large cost
is incurred in the action due to the term
$(\hbar/2e)^2\dot \phi_q(\omega) \dot \phi_q(-\omega)
D_K(\omega)D_R^{-1}(\omega)D_A^{-1}(\omega)$, causing an
attractive interaction between the instanton and
anti-instanton. 
Similar arguments apply for an instanton/anti-instanton pair
on opposite parts of the Keldysh contour.  
The transition between the bound and unbound phase of the instantons
is the diffusion/localisation transition of Ref.\cite{Schmid83}.   A
large value of $\phi_q$  
implies that the system is in a
coherent superposition of very different classical states.
Coupling to a heat bath causes this coherence to decay;
in a sense the heat bath {\it measures} the state of the system.

For an instanton/anti-instanton pair separated by a real time
$\tau$, the quantum component of the junction phase is given by
$\phi_q(t)=2 \pi [ \theta(t+\tau/2)-\theta(t-\tau/2)]$, so that 
$\phi_q(\omega)=\pi \tau \sin (\omega \tau/2)/(\omega \tau/2)$. Using
a high-momentum cut-off, $\Lambda$ and in the limit $T\rightarrow 0$,
$D_K/(D_A D_R)=2 |\omega| (1+2e^{-\hbar |\omega|/kT})/Z$ and the
dissipative contribution to the action is given by
${\cal S}_{Diss}=i(h/2e^2Z)[\ln(\Lambda \tau) +(kT\tau/\hbar)^2]$. 
We interpret $e^{i {\cal S}_{Diss}/\hbar}$ as a distribution function
for $\tau$. At zero temperature, one finds that for $Z<h/2e^2$ the
instantons are bound and tunneling is incoherent. At finite
temperatures, the instantons are, strictly, always bound on a
timescale of order $\tau \sim (2 \pi e/k)\sqrt{2hZ}$; coherence
between tunneling events is lost on timescales greater than this.

Now let us incorporate these ingredients into
an extended Langevin equation. The Langevin equation is
the effective theory of the {\it classical} part of the fields,
represented as a differential equation with noise. 
Of course, it is not always (or
even usually) possible to represent a field theory in this way. 
The key feature above that allowed the
effective theory for the classical components to be written in
Langevin form was that all terms, aside from the coupling to the
bath, could be linearized in the quantum components. 
This required that the quantum components be small, a condition that is
satisfied explicitly in the incoherent hopping regime
\cite{Schmid82,Eckern84,Aashish}.  
In this regime,
we can decouple the quadratic term in $\phi_q$ as before. 
The path integral
reduces to a sum over sectors with different 
configurations of instanton pairs
(we need only consider pairs between
the upper and lower part of the Keldysh contour; those on the same contour will have no effect). 
Each sector is weighted in the
sum by the probability of producing the instanton
pairs\cite{Ford88,Rajaraman}. After integrating out $n_q$ and $\phi_q$
the partition function reduces to 
\begin{eqnarray}
& &
{\cal Z}
=
\sum_{\{ t_a,t_b\}}
\prod_{\{t_a,t_b\}}
\int_{
\tiny{
\begin{array}{c}
\phi_c(t_a-0)
=-\phi_c(t_a+0)
\nonumber
\\
n_c(t_a-0)=-n_c(t_a+0)
\end{array}}}
D\phi_c D n_c
\nonumber\\
& &
\times
\delta 
\left[ 
\left(\frac{\hbar}{2e}\right) \dot \phi- \frac{n}{C} 
\right]
\delta
\left[
\dot n - \left(\frac{\hbar}{2e}\right) \frac{\dot \phi}{Z} + \dot q +2
I_J \sin(\phi)
\right]
\nonumber\\
& &
\times
\exp \left[ -
\frac{\hbar }{\omega Z} \coth(\hbar \omega/2T)q(\omega)
q(-\omega)
\right]
P(t_a)P(t_b)
.
\label{Z1}
\end{eqnarray}
$P(t_a)$ is the probability of producing the instanton pair corresponding to
tunneling and $P(t_b)$ the probability of a
pair corresponding to reflection. The path integrals over
$\phi_c$ and $n_c$ are carried out with the boundary conditions
for the configuration of instantons in that sector.
We encode these boundary conditions
within the $\delta$-functionals as follows:
\begin{eqnarray}
& &{\cal Z}
=
\sum_{\{ t_a,t_b\}}
\prod_{\{t_a,t_b\}}
P(t_a)P(t_b)
\int D\phi_c D n_c
\nonumber\\
& &
\times
\delta 
\left[ 
\left(\frac{\hbar}{2e}\right) \dot \phi- \frac{n}{C} +2
\left(\frac{\hbar}{2e}\right) \sum_a \phi(t_a) \delta(t-t_a) 
\right]
\nonumber\\
& &
\times
\delta
\left[
\dot n - \frac{\hbar \dot \phi}{2eZ} + \dot q +2
I_J \sin(\phi)
+2  \sum_b  n(t_b)\delta(t-t_b)
\right]
\nonumber\\
& &
\times
\exp \left[ -
\frac{\hbar }{\omega Z} \coth(\hbar \omega/2T)q(\omega)
q(-\omega)
\right]
\label{Partition2}
\end{eqnarray}
Eq.~(\ref{Partition2}) is equivalent to the extended Langevin
equation(\ref{Extended_Langevin}). 
The drive current $I_D(t)$ on the right-hand side of the second of
Eqs.(\ref{Extended_Langevin}) arises {\it via} a term
$I_D(t) \phi_q(t)$ in Eq.(\ref{Keldysh_Lagrangian}). The drive leads to a
modification of the instanton action and probability due to the
possibility of photon assisted tunneling\cite{Ivlev85}. 
The extended quantum Langevin equation (\ref{Extended_Langevin}) is
valid in the regime of incoherent hopping provided that the
constituent steps---the derivation of the quantum Langevin equation
itself \cite{Schmid82,Eckern84,Aashish} and the instanton calculation
\cite{Ford88,Ivlev85,Instanton}--- are valid.  

To conclude, we have considered an instanton expansion within the
Keldysh field theory of a Josephson junction coupled to a heat
bath. Using this expansion, we have developed extensions to the
quantum Langevin equation, which incorporate quantum features of the
junction dynamics.  
This extended Langevin equation represents a synthesis of established
techniques for dealing 
with semi-classical dynamics in the presence of environmental
fluctuations\cite{Schmid82,Eckern84} and
tunneling\cite{Ford88,Ivlev85}. It is valid in the regime of
incoherent hopping. 
This extended Langevin equation allows numerical simulation
of the junction dynamics in response to a time-dependent drive
current, including 
quantum and thermal effects of the bath, non-linear classical dynamics
and quantum dynamics of the junction. 

I would like to thank A. Clerk,  S. Girvin, K. Sengupta and I. Smolyarenko
for discussions and assistance. This work was supported by the Royal Society.

\end{narrowtext}

\begin{thebibliography}{99}
\bibitem{Footnote1} Our approach contrasts with that
of Refs\cite{MQT} (which involve expansions in tunneling in a dual,
tight-binding model, or, equivalently, in Bragg scattering in the
Josephson potential) in that it retains the full non-linearity of the
Josephson potential throughout and allows non-linear classical effects
to be described simultaneously. The two approaches are closely
related, since both amount to expansions of the Feynman-Vernon
influence functional\cite{Feynman63}; they agree in all simple limits.
\end{thebibliography}
 \end{document}